\documentclass[twocolumn,showpacs,preprintnumbers,amsmath,amssymb]{revtex4}

\usepackage{graphicx}
\usepackage{dcolumn}
\usepackage{bm}

\newcommand{\he}{$^4$He }
\newcommand{\hee}{$^4$He}
\newcommand{\het}{$^3$He }
\newcommand{\hett}{$^3$He}
\newcommand{\fl}{$^{19}$F }
\newcommand{\fe}{$^{55}$Fe }
\newcommand{\alu}{$^{27}$Al }
\newcommand{\cm}{$^{244}$Cm }
\newcommand{\iso}{C$_4$H$_{10}$ }
\newcommand{\isoo}{C$_4$H$_{10}$}
\newcommand{\nit}{N$_4$S$_3$ }

\def\NIMA#1#2#3{{\rm Nucl.~Instr.~and~Meth.} {\bf{A#1}} (#2) #3}
\def\NIM#1#2#3{{\rm Nucl.~Instr.~and~Meth.} {\bf{#1}} (#2) #3}

\def\PRA#1#2#3{{\rm Phys. Rev.} {\bf{A#1}} (#2) #3}

\def\PRD#1#2#3{{\rm Phys. Rev.} {\bf{D#1}} (#2) #3}

\def\PRL#1#2#3{{\rm Phys.~Rev.~Lett.} {\bf{#1}} (#2) #3}

\def\APJ#1#2#3{{\rm Astrophys.~J.} {\bf{#1}} (#2) #3}

\begin{document}


\title{Ionization Quenching Factor Measurement of \he}
 
\author{D. Santos}
\email{santos@lpsc.in2p3.fr}
\affiliation{ LPSC, Universit\'e Joseph Fourier Grenoble 1,
  CNRS/IN2P3, Institut Polytechnique de Grenoble, Grenoble, France
}

\author{F. Mayet}
\email{mayet@lpsc.in2p3.fr}
\affiliation{ LPSC, Universit\'e Joseph Fourier Grenoble 1,
  CNRS/IN2P3, Institut Polytechnique de Grenoble, Grenoble, France
}

\author{O. Guillaudin}
\email{guillaudin@lpsc.in2p3.fr}
\affiliation{ LPSC, Universit\'e Joseph Fourier Grenoble 1,
  CNRS/IN2P3, Institut Polytechnique de Grenoble, Grenoble, France
}

\author{Th. Lamy}
\affiliation{ LPSC, Universit\'e Joseph Fourier Grenoble 1,
  CNRS/IN2P3, Institut Polytechnique de Grenoble, Grenoble, France
}

\author{S. Ranchon}
\affiliation{ LPSC, Universit\'e Joseph Fourier Grenoble 1,
  CNRS/IN2P3, Institut Polytechnique de Grenoble, Grenoble, France
}
\author{A. Trichet}
\affiliation{ LPSC, Universit\'e Joseph Fourier Grenoble 1,
  CNRS/IN2P3, Institut Polytechnique de Grenoble, Grenoble, France
}
\author{ P. Colas}
\affiliation{ CEA-Saclay, France 
}
\author{ I. Giomataris}
\affiliation{ CEA-Saclay, France 
}

\date{\today}
\begin{abstract}
The ionization quenching factor (IQF) is defined as the fraction of energy released by a recoil in a medium through ionization compared with its total kinetic energy. At low energies, in the range of a few keV, the ionization produced in a medium falls  rapidly and systematic measurements are needed. 
We report measurements carried out at such  low energies as a function of the pressure in \he at 350, 700, 1000 and 1300 mbar. In order to produce a nucleus moving with a controlled energy in the detection volume, we have developed an Electron Cyclotron Resonance Ion Source (ECRIS) coupled to an ionization chamber by a differential pumping.
The quenching factor of \he has been measured for the first time down to 1 keV recoil energies. An important deviation with respect to the phenomenological calculations has been found allowing  an estimation of the scintillation 
produced in \he as a function of pressure. The variation of the IQF as a function of the percentage of isobutane, used as quencher, is also presented.
\end{abstract}

\pacs{Valid PACS appear here}
\maketitle

The measurement of the amount of ionization produced by particles in a medium presents a great interest in several fields from the 
radiation damage and metrology to particle physics and cosmology. Indeed, a nuclear recoil moving at very low energies  can be used to detect rare events such as neutrino coherent interactions or non-baryonic dark matter signatures.
 Cosmological observables such as the temperature anisotropies of the cosmic microwave background 
\cite{wmap,archeops}, the distance of supernov\ae~of type Ia \cite{SNIa,Astier}, the galaxy cluster abundances
\cite{CG} 
and the baryonic acoustic oscillations \cite{BAO} converge to a concordance model \cite{wmap} in which  most of the matter of the Universe should be non-baryonic, in a still unknown nature. 
The direct detection of these non-baryonic particles is based on the detection of nuclear recoils coming from elastic collision in different targets. Ionization is one of the most important channels to detect such nuclear recoils, heat, scintillation and tracking being the others.
The ionization quenching factor (IQF) is the fraction of the kinetic energy released through ionization by a recoil in a medium.
In the last decades an important effort  has been made to measure the IQF in different materials : 
gases \cite{H}, solids \cite{Ge,Si} and liquids \cite{Xe}, using different techniques.  The use of a monoenergetic neutron beam has been explored in solids with success \cite{Qneutron}.
However in the low energy range the measurements are rare or absent for many targets due to ionization threshold  of detectors and experimental constraints. In this letter, we present the results of the 
IQF measurement in \he  using a dedicated experimental setup specially developed at the LPSC in Grenoble to access to the low energy particle range.
The fact that we have succeeded to measure the ionization in \he down to energies of 1 keV recoil opens the possibility to develop a gas detector 
using \het (MIcro-tpc MAtrix of Chambers: MIMAC) to search for the exotic particles presumably composing the local galactic 
dark halo. The use of \het and \fl is motivated by their privileged features for dark matter search. With their odd atomic number, a detector made of such targets will be sensitive to the spin-dependent interaction, leading to a natural complementarity to existing detectors mainly sensitive to scalar interaction \cite{susy}. A large matrix modular gas detector getting the track 
recoil information will be able to correlate the rare events 
with the earth motion with respect to the galactic halo. 
In fact, this directional detection will be the only one to 
validate the existence of a galactic dark halo formed by 
weakly interacting particles (WIMPs)\cite{Drift,MIMAC,Collar,mit}.

The energy released by a particle in a medium produces in an interrelated way three different processes: i) ionization, producing a number of electron - ion pairs, ii) scintillation, producing a number of photons coming from the de-excitation of quasi-molecular states and iii) heat produced essentially by the motion of nuclei and electrons. The way in which the total kinetic energy released is shared between the electrons and nuclei by interactions with the particle  has been estimated theoretically, 
four decades ago \cite{Lindhard}, in very specific cases, those in 
which the particle and the target are the same. Since then, phenomenological studies have been proposed for many (particle,target) systems 
\cite{Hitachi}.
At low energies, in the range of a few keV, the ionization produced in a medium varies very rapidly and systematic measurements are needed.

The measurements reported in this work have been focused on this low energy range with in addition a systematic study of the ionization  as a 
function of the \he gas pressure, one of the most standard gases used in particle detection.
The measurement on \he can be taken as a lower limit for the \het because the phenomenological estimation of the IQF 
\cite{Lindhard,Lewin} for \het is greater than for \he as shown in Fig. 1.

To produce a nucleus moving with a controlled energy in the detection volume, 
we have developed \cite{lamy} an  Electron Cyclotron Resonance Ion Source (ECRIS, \cite{Geller})
with an extraction potential from a fraction of a kV up to 50 kV.
The interface between the ion source and the gas chamber was, in the first series of measurements, a 50 nm thick   
$\rm N_{4}Si_3$ foil. To assure the electrical continuity between the foil and the mechanical 
support 10 nm of Al have been evaporated  to prevent the charging of the foil. A time-of-flight (TOF) measurement has been 
performed  using two channeltrons, one of them detecting the electrons extracted from the foil by the ions and 
the other one detecting the ions at 6 different known distances \cite{Mayetnim}. These TOF set-up allowed us 
to measure the energies of the ions entering the detection volume, just after the foil. 
Having these TOF measurements as an ECRIS output energy reference, we improved the set-up in order to reduce 
the spread of the energy distribution due to the ions straggling inside the 
foil. For this we have removed the foil and installed a $1 \mu m $ hole with a differential pumping. In such a way 
we could verify that the energy values measured by TOF  were the same as those indicated by the potential extraction values in kV for 1+ charge state ions.

The ionization produced in the gas has been measured with a Micromegas (micromesh gaseous) detector 
\cite{Giomataris:1995fq} adapted to a cathode integrated mechanically to the interface of the ECRIS. The Micromegas used was of a type called 
bulk \cite{bulk}, in which the grid and the anode are built and integrated with a fixed gap. This gap depends on the working pressure, being  of 128 $\mu m$ for  measurements between 350 and 1300 mbar. The electric fields for the drift and the avalanche have been selected to optimize the transparence of the grid and the gain for each ion energy. Typical applied voltages were  300 V for the drift and 450 V for the avalanche. The drift distance between the cathode and the grid was  3 cm, large enough to include the tracks 
of \he nuclei of energies up to 50 keV. These tracks, of the order of 6 mm for 50 keV at 1000 mbar, are roughly of the same length as the electron tracks, at the same pressure,  produced by the X-rays emitted by the \fe source used to calibrate. 

Two different calibration sources were used in order to prevent errors coming from the electronic offset. The 1.486 keV X-rays of \alu following the interactions of alpha particles emitted by a source of \cm under a thin aluminium foil  and a standard \fe X-ray source  giving the 5.9 keV K$_{\alpha}$    and the 6.4 keV K$_{\beta}$  lines.
These two lines, as they were not resolved by our detector, have been considered as a single one of  5.97 keV, taking into account their relative intensities. These photoelectron ionization energies provide the calibration needed to get the ionization energies produced by the recoils. The IQF of a recoil will be the ratio between this energy and its total kinetic energy. In such a way, the IQF compares the nuclei ionization efficiency  with respect to the electrons.

The impurities in the gas mixture have been controlled by a  circulating flow keeping the same pressure in the detection volume after a 
vacuum previously obtained of 10$^{-6}$ mbar. This good previous vacuum step was important to prevent the effect of impurities on the W value, the mean energy needed to produce an electron-ion pair. This dependence of W on impurities is well 
known \cite{jesse} and  should be controlled to get the ionization energy values with fluctuations negligible compared to  systematic errors.

\begin{figure} 
\includegraphics[width=7cm,height=8cm,angle=270]{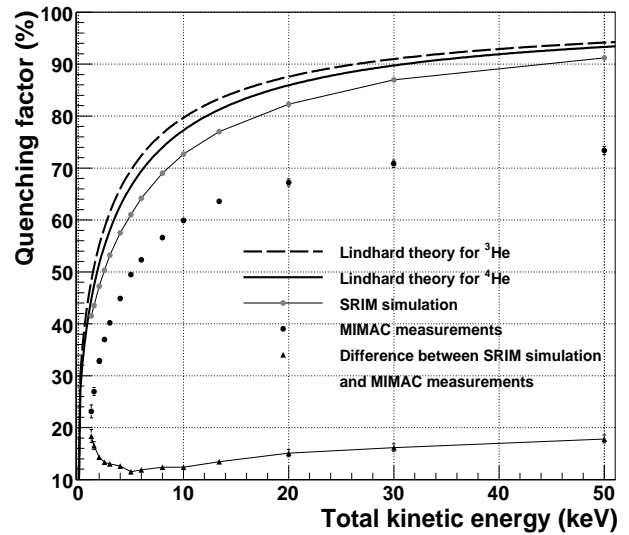}
\caption{
\he ionization quenching factor  as a function of \he total kinetic energy (keV). Lindhard theory prediction (\he in pure \he gas) is presented by a solid line and compared with SRIM simulation results (solid line and points) in the case of \he in \he+5\% \iso mixture. Measured quenching factor are presented at 700 mbar with error bars included mainly dominated by systematic errors. The differences between the SRIM simulation and the measured values are shown by triangles. The Lindhard  
calculation is parametrized 
as in \cite{Lewin}.
The ionization energy calculated with the proper Lindhard parametrization, corresponding to \hett, is always greater than the \he as shown by the dash curve. 
}
\label{plot1}
\end{figure}

\begin{figure}
 \includegraphics[width=8cm,height=8cm,angle=270]{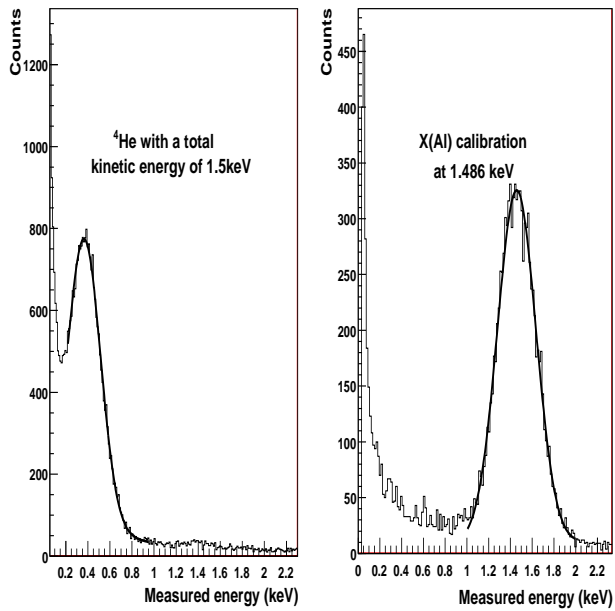}
\caption{Spectra of 1.5 keV total kinetic energy \he (left) and 1.486 keV X-ray of \alu (right) in \he +5\% \iso  mixture at 700 mbar. The comparison of the energy measured between these two spectra gives  the quenching factor of Helium in \he +5\% \iso mixture at 700 mbar. }  
\label{plot2}
\end{figure}

In order to measure the quenching factor of \he in a gas mixture of 95\% of \he and 5\% of 
isobutane (\isoo) we proceeded 
as follows: i) the energy of the ion produced in the ECR source was given by the extraction potential, 
previously checked by the time of flight measurements, as described above, with a thin foil of \nit of 50 nm  
\cite{Mayetnim}, ii) the ionization given by the Micromegas was calibrated 
by the two x-rays (1.486 and 5.97 keV) at each working point of the Micromegas 
defined by the drift voltage (V$_d$), the gain voltage 
(V$_g$) and the pressure, iii) the number of ions per second sent 
was kept lower than 25 pps, to prevent any problem of recombination in 
primary charge collection.\\
Fig. 1 shows the results at 700 mbar compared with the Lindhard theory for \he ions in pure \he and with respect to the SRIM 
\cite{srim} simulation for \he in the gas mixture used during the measurements. The  ionization spectra of \he nuclei  moving in with a kinetic energy of 1.5 keV and of electrons of roughly the same energy (1.486 keV) coming out from  the photoelectric interaction with X-rays of \alu are shown in Fig. 2. Most of these electrons come out from the carbon atomic shells, producing subsequent Auger electrons and fluorescence of 254 eV  maximum energy detected jointly with the photoelectrons. The ratio of these measured energies shown in Fig. 2 is represented as a point in Fig. 1 at 1.5 keV. These quasi-recoil  spectra at very low energies are the result of the coupling of the ion source facility with a Micromegas detector having a very low energy threshold ($\simeq $ 300 eV).
As a control of systematic errors coming from the behavior of the detector, the ionization energy measurements have been made at different gain values, varying (V$_g$) between 390 and 470 V. The fluctuation of the ionization energy values were less than 1\% of the total kinetic recoil energy \cite{Mayetnim}.
We observe a difference between the SRIM simulation and the experimental
points of up to 20\% of the total kinetic energy of the
nuclei, shown in Fig. 1. This difference may be assigned to the scintillation produced by the \he nuclei in \he gas. 
This difference is reduced at lower pressures due to the fact that the amount of scintillation is reduced when the mean distance between the nuclei in the gas is increased giving a lower production probability of  eximer states. 
The production of UV photons in \he by particles moving in as a function of the pressure has been well characterized many years ago 
\cite{chimistes1}. These photons are emitted, besides the discrete lines, mainly in two continuous regions around 67.5 nm and 82 nm 
\cite{chimistes1}. These photons are hard to detect by standard photo detection techniques and require a special experimental setup as used 
in reference \cite{chimistes2}. 

In Fig. 3 the measurements at 350, 700, 1000 and 1300 mbar are shown. We observe a clear, roughly linear, dependence of the IQF on the pressure of the gas that will be reported in a future study down to less than 100 mbar.

\begin{figure}[b]
\includegraphics[width=8cm,height=8.5cm,angle=270]{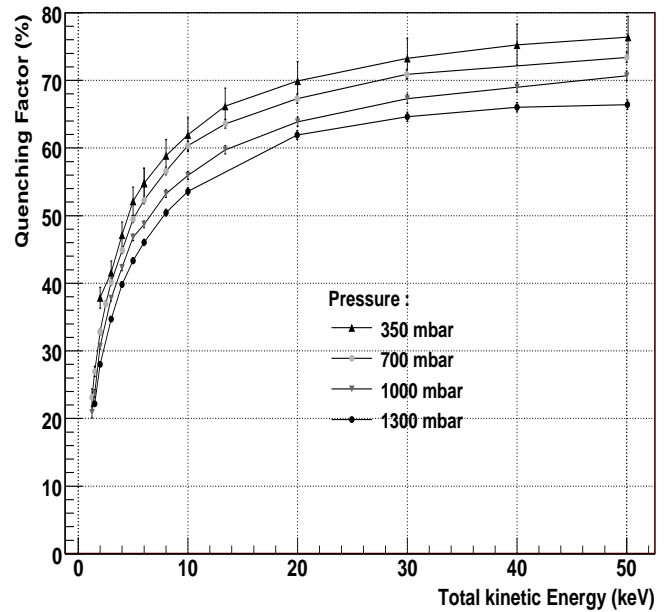}
\caption{ \he quenching factor as a function of \he total kinetic energy (keV) for 350, 700, 1000 and 1300 mbar, in the case of \he in \he +5\% \iso mixture. Measured data are presented with error bars included mainly dominated by systematic errors. Straight line segments between experimental points help to separate the different series of measurements
}  
\label{plot3}
\end{figure}

We performed an additional measurement at 100 keV, extracting at 50 kV  \he$^{2+}$, from the ECRIS, injecting it in a pressure of 1300 mbar to explore the saturation of the IQF at high energies. We got the value of 
$0.68 \pm  0.02 \ \%$ showing a clear saturation unlike  the theoretical prediction that increases asymptotically up to 100\% at high energies. This saturation is probably due to the scintillation production. This fact concerns the particle detection in gases, introducing a non linear response in the energy detection.
The fact that the  amount of ionization increases at low pressures gives an additional chance for the directional direct detection of WIMPs providing even more ionization signal when working at low pressures 
($ \leq $  100 mbar)  having access to the recoil track information. 
In order to get the dependence of the IQF on the percentage of the gas quencher, we performed the measurements at 700 mbar for different percentages of \iso : 2.5\%, 5\%, 7.5\% shown in Fig. 4. The extrapolation of the curve down to 0\% can be taken as an estimation of the ionization quenching factor in 
pure \hee. This variation is another important measurement reported, for the first time, in this work. The quenching factor is highly non linear as a function of the isobutane percentage. Measurements with other quenchers in the future will allow us to compare them and better choose  the gas to optimize the ionization yield.

In summary, we have measured for the first time the ionization quenching factor of \he down to very low energies showing the amount of ionization available to get information of the recoils 
of \hee. The IQF dependence on the pressure of the gas has been described for four different pressures.
An estimation of the scintillation produced in the gas mixture as a function of the energy of the particles has been done. The IQF variation as a function of the quencher has been presented. These measurements are particularly important for searching WIMPs  using \het and in general to better understand the ionization response of helium gas detectors.

We acknowdlege G. Bosson, A. Giganon, F. Lucci, E. Moulin, J. Pancin, A. Pellissier and P. Sortais for their help during the phases of design, development, construction and preliminary data analysis.

\begin{figure}
 \includegraphics[width=8cm,height=8cm,angle=270]{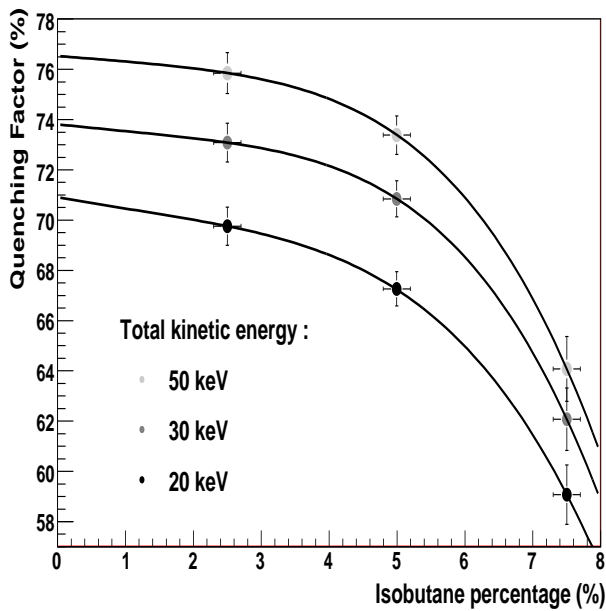}
\caption{ Helium quenching factor  as a function of isobutane fraction  in the gas mixture at 700 mbar, for 20, 30 and 50 keV \he ions. The extrapolation of the fit down to 0\% can be taken as an estimation of the quenching factor of pure Helium. Measured data are presented with statistical and systematic error bars included. }  
\label{plot4}
\end{figure}

\end{document}